\documentstyle[prl,twocolumn,aps,psfig]{revtex}
\input{epsf}

\newcommand{\beq}{\begin{equation}}
\newcommand{\eeqn}[1]{\label{#1}\end{equation}}

\newcommand{\eq}[1]{(\ref{#1})}

\begin{document}

\draft
\twocolumn[\hsize\textwidth\columnwidth\hsize\csname @twocolumnfalse\endcsname

\title{On blowup for Yang-Mills fields}

\author{P.~Bizo\'n$^{1,2}$ and Z.~Tabor$^3$}

\address {$^1$Max-Planck-Institut f\"ur Gravitationphysik, Albert-Einstein-Institut,
 14476 Golm, Germany\\
$^2$Institute of Physics, Jagellonian University, Krak\'ow, Poland\\
$^3$Department of Biophysics, Jagellonian University, Krak\'ow, Poland}

\date{\today}

\maketitle

\begin{abstract}
We study  the development of singularities
 for the spherically symmetric Yang-Mills
equations
in $d+1$ dimensional Minkowski spacetime for $d=4$ (the critical dimension)
and  $d=5$ (the lowest  supercritical dimension).
Using combined numerical and analytical methods we show  in both 
cases that generic solutions starting with sufficiently large initial data blow up in
finite time.
The mechanism of singularity formation depends on the dimension: in $d=5$
the blowup is exactly self-similar while in $d=4$  the blowup is only approximately 
self-similar and 
can be viewed as the
adiabatic shrinking of the marginally stable static solution. The
threshold for
blowup and the connection with critical phenomena in  gravitational
collapse
(which motivated this research) are also briefly discussed.
\end{abstract}

\vskip1pc]

\emph{Introduction:} The Yang-Mills (YM) equations are the basic equations of gauge
 theories describing
 the fundamental forces of nature so understanding their solutions is an issue of
 great importance. This is not an easy task since, 
in contrast to Maxwell's
equations or the Schr\"odinger equation, the YM equations are nonlinear which 
opens up the possibility that
solutions which are initially smooth  become singular in future.
Actually such a spontaneous breakdown of solutions of YM equations cannot occur in 
the physical $3+1$ dimensional Minkowski spacetime as was shown in
a classic 
paper by Eardley and Moncrief~\cite{ea_mo} who proved that solutions
starting from smooth initial data remain smooth for all future times.
A natural question is: how does the property of global regularity depend on the dimension 
of the underlying spacetime, in particular can singularities develop in $d+1$ dimensions
for $d > 3$? We hope that our letter is a step towards answering this question. As we argue below,
the problem of singularity formation for YM equations in higher dimensions is not only interesting in
 its own right but in addition
it sheds some light on our understanding of 
Einstein's
equations in the {\emph{physical}} dimension. 

Despite intensive research the problem of global regularity for YM equations in $4+1$ dimensions
 is entirely open~\cite{kl}.
A lot of progress has been made to prove local existence for "rough" initial data,
yet the attempts of proving global regularity by establishing local well-posedness in the energy norm
fail to achieve the goal by "epsilon"~\cite{kl_ta} (nota bene such a local proof of global 
existence has been obtained  in $3+1$ dimensions~\cite{kl_ma}, thereby improving the theorem of Eardley and
Moncrief).
In this letter we report on numerical simulations which in combination with analytic
results strongly suggest
that generic solutions with 
sufficiently~large energy do, in fact, blow up in finite time. Hence, we~believe~that the above mentioned 
epsilon in the optimal local 
well-posedness result is not a technical shortcoming but is indispensable. We show that the 
 singularity formation is due to  concentration of energy and 
has the form of adiabatic shrinking of the marginally stable static solution. 

Higher dimensions, $d>4$, appear to be somewhat under-explored; the only result we are aware of is the 
proof of existence of self-similar solutions in
$d=5,7,9$~\cite{ca_sh_ta}. These solutions provide examples of singularities developing from smooth initial data,
however nothing was known about their genericity and stability. Here we restrict ourselves to the
$d=5$ case because of its connection with Einstein's equations. We first show that the example of self-similar
 blowup
given in~\cite{ca_sh_ta} is, in fact, generic. Then we look at the threshold for singularity formation and
observe a behaviour similar to the critical behaviour in gravitational collapse~\cite{gu} 
(with blowup being the analogue
of a black hole), in particular we find a self-similar solution with one instability as the critical solution.

We remark in passing that there are close parallels between  YM equations in $d+1$ dimensions and 
 wave
maps in $(d-2)+1$ dimensions~\cite{ca_sh_ta}.
Indeed, many of the phenomena described below have been previously observed by us for the equivariant wave maps
into spheres in two~\cite{bi_ch_ta2} and three~\cite{bi_ch_ta1} spatial dimensions.


\emph{Setup:} We consider Yang-Mills fields in $d+1$ dimensional Minkowski spacetime
 (in the following Latin and Greek
indices take the values $1,2,\dots,d$ and $0,1,2,\dots,d$ respectively). 
The gauge potential $A_{\alpha}$ is
a one-form with values in the Lie algebra $g$ of a compact Lie group $G$. 
Here we take $G=SO(d)$ so the elements of $g=so(d)$ 
can be considered as skew-symmetric $d\times d$ matrices and the Lie bracket is
the usual commutator. In terms of the curvature
$F_{\alpha\beta}=\partial_{\alpha}A_{\beta}-\partial_{\beta}A_{\alpha}+e 
[A_{\alpha},A_{\beta}]$ the
Yang-Mills equations are
\beq
\partial_{\alpha}F^{\alpha\beta}+ e [A_{\alpha}, F^{\alpha\beta}] = 0,
\eeqn{ymeq}
where $e$ is the gauge coupling constant. It is customary to set $e=1$ and we shall
also  do so in the
following. However, it is worth remembering that 
$[e^2]=M^{-1} L^{d-4}$ (in $c=1$ units); in particular  
$e^2$ has the 
same dimension in  $d=5$ as
Newton's constant $G$ in $d=3$.

The YM equations \eq{ymeq} are scale  invariant: if $A_{\alpha}(x)$ is a solution, 
so is
$\tilde A_{\alpha}(x)=\lambda^{-1} A_{\alpha}(x/\lambda)$. The conserved energy
\beq
{\mathcal{E}}(A) = \int_{R^d} Tr\left(F_{0i}^2 + F_{ij}^2\right) d^d x
\eeqn{energy}
scales as
${\mathcal{E}}(\tilde A)=\lambda^{d-4} {\mathcal{E}}(A)$, thus in the PDE terminology the YM equations 
are subcritical for
$d\leq 3$, critical for $d=4$, and supercritical for $d\geq 5$. It is believed that
subcritical equations are globally regular because  energy conservation rules out concentration
of solutions on arbitrarily small scales. In contrast, for supercritical equations
concentration might be  energetically 
favourable  and consequently singularities are expected to occur. The critical equations
provide an interesting borderline case.

We assume  the spherically symmetric ansatz~\cite{du}
\beq
A^{ij}_{\alpha}(x) = \left(\delta^i_{\alpha}x^j-\delta^j_{\alpha}x^i\right)
\frac{1-w(t,r)}{r^2},
\eeqn{ansatz}
where $r=\sqrt{x_i^2}$. Then,
the YM equations \eq{ymeq} reduce to the scalar semilinear wave 
equation for the magnetic gauge potential $w(t,r)$
\beq
w_{tt}= \Delta_{(d-2)} w  + \frac{d-2}{r^2} w (1-w^2) =0,
\eeqn{eqn}
where $\Delta_{(d-2)}=\partial_r^2+\frac{d-3}{r}\partial_r$ is the radial Laplacian 
in $d-2$ dimensions. We solve numerically the initial value problem
for the above equation in $d=4$ and $5$. Our simulations were performed using
finite-difference methods combined with adaptive mesh refinement. The latter was essential is
resolving the structure of singularities developing on very small scales.
To ensure
regularity at the center we require that $w(t,0)=1+O(r^2)$.  At the outer boundary of the
 computational grid we impose the outgoing wave condition.
Below we present results for the time-symmetric gaussian initial data of the form
\beq
w(0,r)= 1- A r^2 \exp\left[- \sigma (r-R)^2\right], \quad w_t(0,r)=0,
\eeqn{data}
with adjustable amplitude $A$ and fixed parameters $\sigma=10$ and $R=2$. 
We have obtained the same qualitative results for several other
families of initial data so we believe that the phenomena described here are generic.

\emph{Results:} We begin our description in a unified
 dimension-independent manner; all statements which do not explicitely involve the dimension 
 apply both to $d=4$ and $d=5$.
 Since our data are time-symmetric, the initial profile splits into ingoing and outgoing waves.
 The evolution of the outgoing wave has nothing to do with singularity formation so we shall ignore it in what
 follows. The behaviour of the ingoing wave depends on the amplitude $A$. 
For small amplitudes the ingoing wave approaches the center, reaches a minimal radius, bounces back
and then disperses to infinity leaving behind an empty space. 
For large amplitudes the ingoing wave keeps concentrating near the center
and eventually blows up in finite time. As the blowup time $T$ is approached we observe
the development of a rapidly evolving inner
region  which is clearly separated from an almost frozen outer region. 
The inner solution attains 
a  kink-like shape which shrinks in a self-similar manner
\beq
w(t,r) \approx W(\eta), \quad \eta=\frac{r}{\lambda(t)},
\eeqn{inner}
where the profile $W$ depends on the dimension but otherwise seems universal.
The scale $\lambda(t)$ goes to zero as $t \rightarrow T$ which signals blowup since
the second derivative $\partial^2_r w(t,0)=W''(0) \lambda^{-2}(t)$
becomes unbounded (there is no blowup of the first derivative because $W'(0)=0$, as we shall see below).

The subsequent discussion of the details of blowup has to be given separately in each dimension.
We begin with the easier supercritical case.\vskip 0.1cm
\noindent $d=5$: In this case the scale changes linearly, that is  $\lambda(t)=T-t$, as one would expect from dimensional
analysis. The blowup profile is given by the exact self-similar solution of equation \eq{eqn}
\beq
W=W_0(\eta)=\frac{1-\eta^2}{1+\frac{3}{5}\eta^2}, \quad \eta=\frac{r}{T-t}.
\eeqn{css0}
\vspace{-0.4cm}
\begin{figure}
\epsfxsize=8.6cm
\epsffile{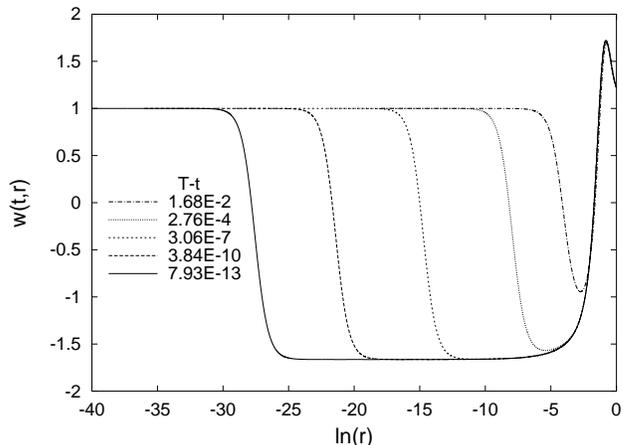}
\vskip 0.1cm
\caption[]{Blowup in $d=5$. The plot shows the late time evolution of large initial data ($A=0.2$). 
As the blowup progresses, the inner solution gradually attains
 the form of the stable self-similar solution $W_0(r/(T-t))$. The outer solution appears frozen
on this timescale.}
\label{fig1}
\end{figure}
\noindent The solution $W_0$ was proved to exist in~\cite{ca_sh_ta} and recently found in closed form by one of
us~\cite{bi}. This solution is linearly stable~\cite{bi} (apart from the
instability corresponding to shifting the blowup time) which supports the fact that we see it as a generic
attractor without tuning any parameters of initial data (see Fig.~1).

 We think that the basic mechanism
which is responsible for the observed asymptotic self-similarity of blowup can be viewed as the convergence to
the lowest "energy" configuration. To see this, note that rewriting \eq{eqn} in terms of the similarity
variable $\eta$ and the slow time $\tau=-\ln(T-t)$ one can  convert the problem of blowup into the 
problem of asymptotic behaviour of solutions for $\tau \rightarrow \infty$. The point is that in these
variables
the wave equation contains a damping term which simply reflects the presence of an outward flux of energy 
through the past light cone of the singularity. Hence it is natural to expect that  solutions will tend
asymptotically to the least "energy" equilibrium state, which is nothing else but $W_0$. 

It was shown in~\cite{bi} that  $W_0$ is actually the ground state of a countable  family of self-similar 
solutions $W_n$ ($n=0,1,\dots$) of equation \eq{eqn}. All $n>0$  solutions are unstable and therefore
 not observed in the evolution of generic initial data. However, they may show up in the evolution of specially
 prepared initial data.
The solution $W_1$ with one unstable mode is particularly interesting since it appears as a transient metastable 
state in the evolution of initial data tuned to the threshold for blowup. 
This indicates that $W_1$ is a critical solution whose codimension-one stable manifold separates blowup from 
dispersion (see Fig.~2).
\vspace{-0.1cm}
\begin{figure}
\epsfxsize=8.6cm
\epsffile{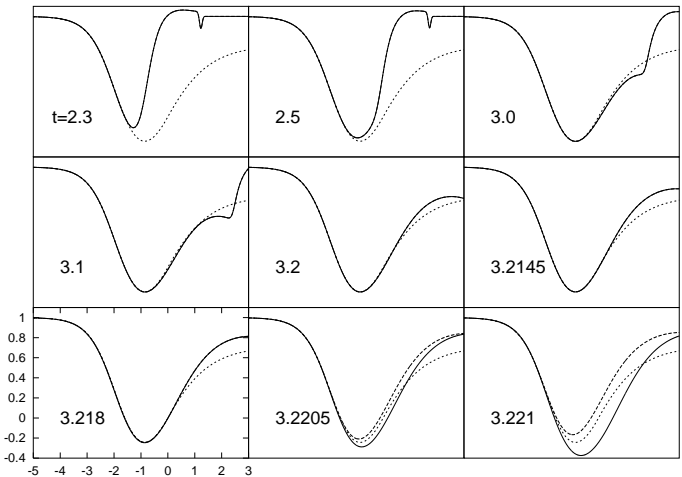}
\vskip 0.5cm
\caption[]{The critical behaviour in $d=5$. The rescaled solution $w(t,(T-t)r)$ 
is plotted against $\ln(r)$
for a sequence~of intermediate times. 
Shown (solid and dashed lines) is  the pair of solutions starting with marginally critical
 amplitudes $A=A^*\pm \epsilon$, where $A^*=0.144296087005405$. Since 
$\epsilon=10^{-15}$, the two solutions are  indistinguishable on the first seven frames. The convergence to the 
self-similar solution $W_1$ (dotted line) is clearly seen
in the intermediate asymptotics. 
The last two frames show the solutions departing from the intermediate attractor towards blowup and dispersion,
respectively. 
}
\label{fig2}
\end{figure}
 Let
$A^*$ be the critical amplitude corresponding to the threshold. For initial data with amplitudes near
 $A^*$, in the intermediate asymptotics when the solution hangs around $W_1$, 
the amplitude of the unstable mode about $W_1$ is proportional to 
$(A^*-A) (T-t)^{-\gamma}$ where $\gamma=5$ (!) is the eigenvalue of the unstable mode~\cite{bi}.
This implies that the time of departure from the intermediate attractor, call it $t^*$, scales as
$T-t^* \sim |A^*-A|^{1/5}$. 
Various scaling laws can be derived from this. For example, consider solutions with marginally 
subcritical amplitudes $A=A^*-\epsilon$. For such solutions the energy density
\beq
\rho(t,r)= \frac{w_t^2}{r^2}+ \frac{w_r^2}{r^2}+ \frac{3(1-w^2)^2}{2 r^4}
\eeqn{density}
initially grows at the center, attains a maximum at a certain time $t_{max}$ and then drops to zero.
 An elementary dimensional
analysis based on the above scaling predicts the power law 
$\rho(t_{max},0)] \sim \epsilon^{-4/5}$. We have verified this prediction 
 in our simulations (with $4\%$ error).

We  point out that the threshold behaviour described above shares many features
with  critical phenomena at the threshold for black hole formation in  gravitational collapse~\cite{gu}.
This fact, together with similar results for wave maps in $3+1$ dimensions~\cite{bi_ch_ta1},~\cite{li_hi_is},
(and other systems~\cite{br}) 
shows that the basic properties of critical collapse, such as universality, scaling, and self-similarity, 
originally
observed for Einstein's equations, actually have nothing to do with gravity and seem to be robust
properties of supercritical evolutionary PDEs.
\vskip 0.1cm
\noindent $d=4$: In this case equation \eq{eqn} does not admit regular self-similar solutions,
so the numerically observed self-similarity can be only approximate. We identify the blowup profile
as the scale-evolving \emph{static} solution (see Fig.~3)
\beq
W=W_S(\eta)=\frac{1-\eta^2}{1+\eta^2}, \quad \eta=\frac{r}{\lambda(t)}.
\eeqn{instanton}
We call this solution static because for any \emph{fixed} $\lambda$ it is the time-independent solution of equation
\eq{eqn} (this solution is perhaps better known  as the YM instanton in four euclidean dimensions). 
Since the energy does not depend on $\lambda$, these solutions
 are only marginally stable: when kicked they shrink or expand. 
In other words the blowup can be viewed as the adiabatic shrinking of the static solution.
Numerical evidence suggests that the rate of  blowup goes asymptotically to zero, that is ($\;\dot{}=d/dt$)
\beq
 \lim_{t\rightarrow T} \frac{\lambda(t)}{T-t} = -\lim_{t\rightarrow T} \dot \lambda =0.
  \eeqn{rate}
  Although we are not able to explain this fact, we point out that it seems necessary for the consistency 
  of the quasi-static character of blowup \eq{instanton}.
 To see this, 
 substitute \eq{inner} into
\eq{eqn} to obtain 
\begin{eqnarray} \label{approx}
 \nonumber (1- \dot \lambda^2 \eta^2) W''  &+ &\left[1+ (\lambda
\ddot \lambda -2\dot \lambda^2) \eta^2\right] \displaystyle{\frac{W'}{\eta}}\\
 &+&\displaystyle{\frac{2}{\eta^2}} W (1-W^2) =0.
\end{eqnarray}
It follows from \eq{rate} that the terms involving time derivatives of $\lambda$ in \eq{approx}
become asymptotically negligible and therefore in the leading order this equation has the same form 
as the right-hand side of \eq{eqn}, which explains why the blowup profile has the shape of the static 
solution.

 In view of the limited resolution of our numerics
 and the lack of theory we are not in position to make conjectures about the time dependence of $\lambda$
 which would go beyond equation \eq{rate}. 
  Although a power law
fit $\lambda \sim (T-t)^{1+\alpha}$ with the anomalous exponent $\alpha\approx 0.1$ is quite 
accurate, we would not take this fact too seriously because we cannot rule out logarithmic corrections
and, moreover, the exponent $\alpha$ exhibits weak dependence on initial data.
\vspace{-0.2cm}
\begin{figure}
\epsfxsize=8.6cm
\epsffile{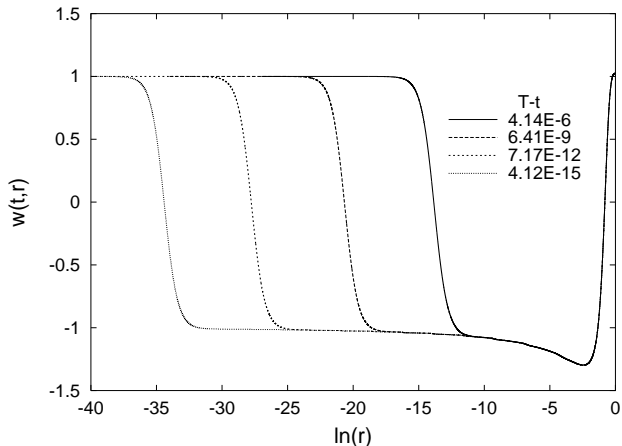}
\vskip 0.09cm
\caption[]{Blowup in $d=4$. The plot shows the late time   evolution of large initial data ($A=0.5$).
 The inner solution has the form of the scale-evolving static solution
$W_S(r/\lambda(t))$ with the scale $\lambda(t)$ going to zero slightly faster than linearly.}
\label{fig3}
\end{figure}
\vspace{-0.1cm}
It is instructive to compare the $d=4$ and $d=5$ blowups from the standpoint 
 of energy concentration.
To this end, for solutions which  blowup we define  the energy  at time $t<T$ inside 
the past light cone of the singularity
\beq
{\mathcal E}(t) = c(d) \int_0^{T-t} \left(w_t^2 +w_r^2 +\frac{d-2}{2 r^2} (1-w^2)^2\right)  
r^{d-3} dr,
\eeqn{energy_cone}
where the coefficient $c(d)=(d-1) vol(S^{d-1})$ follows from integrating \eq{energy} over the angles and taking the trace.
For $d=5$, substituting \eq{css0} into \eq{energy_cone} we obtain $\lim_{t \rightarrow T} {\mathcal E}(t)=0$,
hence no energy gets concentrated into the singularity. In contrast, for $d=4$, 
substituting \eq{instanton} into \eq{energy_cone} and using \eq{rate} we obtain
$$
\lim_{t \rightarrow T} {\mathcal E}(t)= 6 \pi^2 \int_0^{\infty} \left({W'_S}^2 + \frac{(1-W_S^2)^2}{r^2} \right) r dr 
= 16 \pi^2,
$$
thus the energy equal to the energy of the static solution $W_S$ concentrates at the singularity.
Note that only the potential energy becomes concentrated; the kinetic energy tends asymptotically to zero.
In fact, in our simulations we can see the excess energy being slowly radiated away from the inner region as the blowup
profile converges to the static solution.

To summarize, our work provides numerical evidence that solutions of YM equations in four and five spatial
 dimensions
do form singularities from generic smooth large initial data. While the self-similar character of blowup
in $d=5$ is well understood (at least from the numerical perspective), the $d=4$ case is more subtle and our
analysis leaves two important questions open, namely: what is the precise rate of blowup and what is the nature
of the threshold for blowup? We plan to approach these issues by interpreting
the expression \eq{instanton} in terms of motion along the one-dimensional moduli
space of static solutions with the scale $\lambda$ playing the role of the collective coordinate.
The most straightforward way of computing the dynamics on the moduli space using the geodesic
approximation is too naive in the present case: it predicts that $\lambda$ changes linearly with time~\cite{li}
which  contradicts  our numerics. We hope that more refined methods of dealing with
collective coordinates,
like  the ones described in~\cite{fi_pa} in the context of the nonlinear Schr\"odinger
 equation in two spatial dimensions, can be applied to our problem as well. In our opinion,
 the derivation of the correct modulation equation for the scale $\lambda$ 
 is the most important next step towards understanding
the dynamics of blowup for  YM equations in four spatial dimensions; 
hopefully it would  also shed light on  the character of 
transition between blowup and dispersion.

\emph{Acknowledgment.} This work was supported in part by the
 KBN grant 2 P03B 010 16.
  We thank Tadek Chmaj for collaboration during the early stage of our research.

\end{document}